# Big data and big values: When companies need to rethink themselves

Barchiesi, M. A., & Fronzetti Colladon, A.

# Big data and big values: When companies need to rethink themselves

Barchiesi, M. A., & Fronzetti Colladon, A.

**Abstract**

In order to face the complexity of business environments and detect priorities while triggering contingency strategies, we propose a new methodological approach that combines text mining, social network and big data analytics, with the assessment of stakeholders' attitudes towards company core values. This approach was applied in a case study where we considered the Twitter discourse about core values in Italy. We collected more than 94,000 tweets related to the core values of the firms listed in Fortune's ranking of the World's Most Admired Companies (2013-2017). For the Italian scenario, we found three predominant core values orientations (Customers, Employees and Excellence) – which should be at the basis of any business strategy – and three latent ones (Economic/Financial Growth, Citizenship and Social Responsibility), which need periodic attention. Our contribution is mostly methodological and extends the research on text mining and on online big data analytics applied in complex business contexts.

# 1. Introduction

The business scenarios in which companies cooperate and compete nowadays are exponentially complex (Ferraro & Iovanella, 2015; Ferraro, Iovanella, & Cinelli, 2017). Complex decisions are in the daily practice of top business managers and, when successful, they often leverage on the discovery and acquisition of new knowledge from multiple information sources. In the past decades, the emergence of big data and related analytics has drastically increased the amount, variety, and access velocity of the data which companies can collect and manage in their decision making processes. While data integration and the technological aspects of big data were largely discussed in the work of both academics and practitioners (De Mauro, Greco, & Grimaldi, 2016; Yaqoob et al., 2016), the role and the impact of complexity tools in redesigning company decision-making processes, management practices and competitive strategy formulation, still remain partially unexplored. This is for a part surprising, given the importance that big data analytics can have on management decisions (Janssen, van der Voort, & Wahyudi, 2017; Merendino et al., 2018), supporting, for instance, a better organizational design (Schildt, 2017). To achieve a better performance in complex environments, big data must be leveraged appropriately by firms, which need to develop distinctive and difficult-to-imitate capabilities (Gupta & George, 2016; Mikalef, Framnes, Danielsen, Krogstie, & Olsen, 2017; Mikalef, Ilias, Giannakos, Krogstie, & Lekakos, 2016; Pavlou & El Sawy, 2006; Wang, Kung, Wang, & Cegielski, 2017). Data acquisition investments alone are not sufficient to generate business value: firms need to develop a digital business strategy (Bharadwaj, El Sawy, Pavlou, & Venkatraman, 2013) - i.e. an organizational strategy formulated and implemented by leveraging digital resources to create differential value. An effective strategy must be technically practicable, politically acceptable to stakeholders, and has to accord with an organization's culture and core values (Bryson, Crosby, & Bryson, 2009). It is generally recognized in the strategic literature that if a company wants to successfully execute a strategy, it needs to align its organizational culture, core values, systems and processes, resources and capabilities with that strategy (Higgins & Mcallaster, 2004; Waterman, 2007).

The methodology presented and discussed in this paper aims to analyze the stakeholders' attitude towards company core values, in order to help managers better addressing the development and execution of digital business strategies. Indeed, core values influence the strategy performance, as they lie at the interface between internal and external stakeholders (Yoganathan, McLeay, Osburg, & Hart, 2017). Employees - inspired by the organizational core values - translate an organization's strategy into reality through the complex interactions with other stakeholders (Ind, 2004). Over time, firm's organizational core values should evolve, under the pressure of the different tensions arising from the diverse point of views and competing interest amongst stakeholders (Urde, 2009; Yoganathan et al., 2017). Indeed, core values which exist only in the organizational rhetoric and which are meaningless to stakeholders are to be considered as hollow (Urde, 2009), as they will be hardly translated into competitive advantage and commercial success.

Even if many organizations make an effort to define their core values and emphasize them in their internal and external communication activities, they rarely use structured methodology, based on big data and complexity tools, to evaluate how these values are perceived by stakeholders.

Nowadays, social media websites express the voices and thoughts of a large variety of users worldwide, and the vast majority of companies recognize their potentials, also to determine their competitive position (Pappas, Mikalef, Giannakos, Krogstie, & Lekakos, 2017). Indeed, social media transformed the way companies relate to stakeholders (Rybalko & Seltzer, 2010). Platforms like Facebook, TripAdvisor and Twitter can be used by firms to track and measure the outcomes of competitive strategies - even of those which demand significant and complex organizational changes (Xie, Wu, Xiao, & Hu, 2016), as rethinking organizational culture and core values. Following this lead, we propose a methodology for a continuous monitoring system of the stakeholders' attitude towards company core values, based on big data analytics techniques applied to social media. The approach we propose is based on the framework of analysis introduced by Gloor and colleagues (Peter Gloor, Fronzetti Colladon, Giacomelli, Saran, & Grippa, 2017), which overcomes some of the limitations of previous works only focused on the study of one aspect of online communication – such as users' sentiment. As illustrated in Section 3, the framework combines social network and semantic analysis and considers all the three dimensions of online social interaction: structure, use of language and interactivity. The proposed methodology has been tested on Twitter: over a period of two months we collected more than 94,000 tweets representative of the discourse about companies' core values in Italy.

The methodology and the study we present in this paper could be replicated, in several different contexts, to define a contingent core value hierarchy, useful to improve digital business strategies and internal decision making processes. On the one side, we extend the research about the use of big data to support the formulation and implementation of contingency business strategy, on the other, we present a new framework of analysis and a methodology which can be implemented by companies to evaluate the stakeholders' perception about their core values.

## 2. Theoretical Background

Over the past decades, the digital technologies reshaped traditional business strategies, processes, firms' capabilities (Banker, Hu, Luftman, & Pavlou, 2010; Ettlie & Pavlou, 2006; Kohli & Grover, 2008; Sambamurthy, Bharadwaj, & Grover, 2003) and relationships with customers and stakeholders (Ngai, Xiu, & Chau, 2009; Tajvidi, Richard, Wang, & Hajli, 2018) . Social media became essential instruments of dialogue with stakeholders and important analytical sources, to extract consumers' insights and predict sales (Arias, Arratia, & Xuriguera, 2014; Culnan, McHugh, & Zubillaga, 2010; Felix, Rauschnabel, & Hinsch, 2017). The integration of newly developed digital systems with traditional business practices creates a new challenge for enterprises. According to Bharadwaj et al. (Bharadwaj et al., 2013), to fully exploit the potentialities of this integration, it is necessary to rethink the role of IT strategy from that of a functional-level strategy, subordinate to business strategy, to one that is a synthesis between IT strategy and business strategy – which they termed digital business strategy (DBS). The DBS is, at a practical level, an organizational strategy formulated and executed by leveraging digital resources to create differential value. DBS is not just an evolution of the IT strategy, but a more comprehensive strategy that concerns both strategic decisions and other aspects of a company's life, thus not only affecting the IT functional area. Indeed, DBS is different from the traditional IT strategy because: (1) it highlights the pervasiveness of the digital resources in all functional areas; (2) it recognizes digital systems and technologies as resources; (3) it is directly linked to value creation. In addition, DBS formulation and implementation requires a systemic perspective to consider together structures, systems, leadership behavior, human resource policies, culture, values, and management processes (Beer & Eisenstat, 2000; Pryor, Anderson, Toombs, & Humphreys, 2007). Introducing new strategies implies changes along these major strategy execution factors (Higgins & Mcallaster, 2004). It is very difficult to change an organization's culture totally. However, digital resources, together with text mining and social network analysis, may support a strategical shift towards values and norms which can be more synergic with stakeholders' perceptions. In particular, big data and the related analytics can be used by firms to implement continuous monitoring systems of stakeholders' attitudes towards organizational core values. Being aware of the requests of a wide range of stakeholders is at the basis of better long-term strategies, as it compels firms to carefully consider factors and variables that would otherwise be neglected, but that in the long run may result very influential on the business (Barchiesi & La Bella, 2014). Implementing business strategies and managerial practices incompatible with the most critical values of stakeholders can have devastating repercussions on the achievement and sustainability of a competitive advantage (Pekka, 2010).

Indeed, organizational core values orientate organizations' perspective regarding strategic direction, mission determination, and visioning (Barchiesi & La Bella, 2014). They are relatively stable over the time, but not immutable (Yoganathan et al., 2017). Core values are the principles and criteria influencing individual and group behaviors. They are the most defining characteristics of an institution (Williams, 2002), "collective beliefs about what entire enterprise stands for, takes pride in, and holds of intrinsic worth" (Fitzgerald & Desjardins, 2004) . Notwithstanding these fundamental aspects, organizational core values may differ in terms of their desirability over time and space (Gimeno, Labadie, Saris, & Mendoza, 2006; Parada, Nordqvist, & Gimeno, 2010; Yoganathan et al., 2017); the process of changing values is sometimes necessary and always complex and highly strategic, as it involves all the functional areas of an organization. Companies who plan such a comprehensive changing process cannot underestimate the information coming from the analysis of digital data sources. In this context, social networking platforms offer the opportunity to reach a wide range of stakeholders and to capture their feelings and perceptions (Lovejoy, Waters, & Saxton, 2012; Sedereviciute & Valentini, 2011).

## 2.1. Core value orientations

To implement a continuous assessing system of the stakeholders' attitude toward companies' core values, and to define a core value hierarchy, it is useful to introduce the concept of organizational core value orientation.

Organizational values are narrowly related to the organizational culture (Agle & Caldwell, 1999; Chatman, 1989; Vandenberghe & Peiro, 1999) and address the ethical behavior of individuals at work (Akaah & Lund, 1994; Singhapakdi & Vitell, 1993). Accordingly, organizational core value orientation was defined as the representation of the nature of beliefs and thoughts which are subtended in core value expressions (Barchiesi & La Bella, 2014). For example, expressions such as "Leaders start with the customer and work backwards. They work vigorously to earn and keep customer trust...", "[…] we respect your privacy and protect it with strong encryption, plus strict policies that govern how all data is handled.", and "Measuring our success by that of our customers… always driven by Six Sigma quality and a spirit of innovation" denote values that may induce different strategic perspectives and different managerial decisions, but all of them share the same nature: Customers orientation. Following this approach, a set of five core value orientations can be identified (Barchiesi & La Bella, 2014): customers/users, employees, economic and financial growth, excellence, and corporate social responsibility. Although in literature there are several categorizations of core values (Agle & Caldwell, 1999; Cording, Harrison, Hoskisson, & Jonsen, 2014; Hitlin & Piliavin, 2004), we have adopted the one of Barchiesi and La Bella (2014) because it allows to draw conclusions on stakeholders' perception about different areas of the business activity. We briefly recall their categorization in the following, proposing a reorganization of the last orientation, dividing it into two different parts.

*Customers/users orientation*. All the core values related to a customers/users-oriented organizational culture, in which customer satisfaction is at the core of every decision, are defined "customer/user oriented". Expressions of corporate core values that may be included in this area are: "focus on the user and all else will follow", "quality", "service", "passion for our customers", and so on.

*Employees orientation*. In this area, there are all the core values related to a human-resources-oriented organizational culture, in which each employee is seen as a source of strategic competitive advantage. Expressions of corporate core values that can be included in this area are: "[We] recognize exceptional talent, and willingly move them throughout the organization", "Creating a culture of warmth and belonging", "team spirit", "safety and health at work", "a fun and rewarding place to work", and so on.

*Economic and financial growth orientation*. This orientation comprises all the core values which pertain to an organizational culture where economic prosperity and financial strength are considered key factors to be sustained over time, by pursuing stakeholders' interests and maintaining competitive advantages. Expressions of corporate core values that may be included in this area are: "superior financial returns to our shareholders", "attractiveness for investors," and "we grow our business profitably".

*Excellence orientation*. When core values indicate a culture in which all employees' intents and actions are focused on the attainment of outstanding results, the organization is "oriented to excellence". Expressions of corporate core values that might be included in this area are: "hire and develop the best", "think big", "a will to win," "leadership," "embracing speed and excellence" and "great just isn't good enough".

*Corporate social responsibility orientation*. This orientation comprises core values which inspire honest and ethical behaviors in response to social demands, also regarding environmental sustainability and global relevance of the impacts of organizational actions. In our study, we propose to divide this area into two distinct parts: the first, which we call *Citizenship Orientation*, includes all the core values concerned with ethics, such as "integrity", "obey the law", "respect our social and physical environment", "you can make money without doing evil" and so on; the second, which we call *Social Responsibility Orientation*, concerns all the core values expressing the company's contribution to the improvement of society and environment, such as "serve our world", "support agriculture and rural development", "nutrition, health, wellness", "build social value" and so on. We maintain the usefulness of this diversification because it distinguishes the companies' voluntary contribution to the improvement of society and environment (Social Responsibility area), from the correct behavior which companies have to keep because of the shared values, norms and practices of the system in which they operate (Citizenship area).

## 3. Methodology

### 3.1. Study sample

Our case study was based on the analysis of the microblogging platform Twitter, which potentially captures the dialogue of a large set of users, who are effective or potential companies' stakeholders, such as financial traders, politicians, other companies, activists, news outlets, or customers. This platform delivers a fragmented mix of information in almost real time, generated by a large number of users. Twitter has drawn the attention of many scholars and it proved to effectively reflect the collective consciousness for several real world phenomena (P. A. Gloor, Fronzetti Colladon, Miller, & Pellegrini, 2016); it has been frequently used for numerous purposes, such as forecasting stock market prices (Zhang, Fuehres, & Gloor, 2011), or assessing value creation of companies on social media (Culnan et al., 2010). This social platform has also shown to be a major communication channel of companies with stakeholders (Mamic & Almaraz, 2013; Rybalko & Seltzer, 2010). In this study, we crawled Twitter extracting the discourse about the core values of the most admired international business companies[1]. In order to define our search queries, we started from the analysis of the core value statements of the companies which were included for five consecutive years in Fortune's ranking of the World's Most Admired Companies[1], from 2013 to 2017. Subsequently, we extracted the keywords that could better represent the meaning expressed by each statement (later extended with their synonyms). With this approach, we were able to collect more than 94,000 tweets, posted from more than 36,000 users, over a period of two months in 2017. During this period there were no special events involving the companies we analyzed – such as new product announcements or scandals involving firms. In a second step, we filtered out about 30% of the collected tweets, since they were not relevant to the business context.

---

[1] http://fortune.com/worlds-most-admired-companies/

## 3.2. Analytical framework

Maintaining the importance of reading together the signals coming from the social interactions and from text mining (P. A. Gloor, 2017; Peter Gloor et al., 2017; Wasserman & Faust, 1994), we combined methods and tools from semantic and social network analysis.

The semantic and social network metrics we used in our study are described in the following sections and are consistent with the framework presented in the work of Gloor et al. (Peter Gloor et al., 2017), which suggests analyzing social interactions considering the three dimensions of interactivity, connectivity and use of language. This framework is based on the measurement of honest signals of communication and collaboration (P. A. Gloor, 2017; Pentland, 2008), which can reveal the interest of social media users for a certain topic, their sentiment and emotionality, their shared participation and interaction dynamics, and their commitment to online conversations.

### 3.2.1. Degree of connectivity

The degree of connectivity was studied considering the metrics of *Density, Group Betweenness Centrality* and *Group Degree Centrality*. The last two metrics, extensively discussed in the work of Wasserman and Faust (Wasserman & Faust, 1994), represent the extent to which the network is reliant on few central nodes who dominate most of the interactions, or the alternative case where central positions are more equally distributed among social actors. Both measures vary in the range [0, 1], where 1 expresses a maximum centralization. Specifically, group degree centrality takes into account the degree of nodes (i.e. the number of arcs that originate or terminate at each node) and reaches its maximum value in a star graph. In such a case, the network is very centralized, as the central node can reach all its peers with a direct contact, whereas the other peers cannot. The same network configuration would produce a maximum value for the group betweenness centrality, a measure calculated considering the variability in individual betweenness centrality scores of social actors. Betweenness centrality is higher when a node more frequently lies in the shortest paths that interconnect all the other in the network. Accordingly, if every other node has to pass by a central node to interact with the others, this central node will have all the connecting power; on the other hand, if more nodes share a central position, then the positional power is more shared.

*Density* represents the extent to which network nodes are densely connected with a high number of links: this measure is calculated as the ratio of actual connections over the total number of potential connections. The density value has to be interpreted together with the number of actors, as it is probable that larger networks are sparser.

Overall, the degree of connectivity represents the structural dimension of our analysis (Peter Gloor et al., 2017). It is higher when a network is densely connected and participated more broadly – which usually translates into average/low values of group degree and betweenness centrality and a network denser than those totally centralized.

### 3.2.2. Degree of interactivity

In order to study network interactivity, we started measuring the network *Activity* as the number of tweets, retweets, users' mentions and answers, during a specific timeframe. This measure corresponds to the number of network links and proved to be more informative than the simple count of tweets (Zhang et al., 2011). Peaks in the activity can be a signal of specific events involving for instance a company. Subsequently, we considered the *Average Response Time (ART)* taken by users to answer to a tweet (or mention) directed to him/her. Together with this measure, we also considered the one of *Nudges*, which counts the number of pings a user has to receive before interacting with a peer who solicited him with a mention or an answer. Lastly, we calculated *Group Betweenness Centrality Oscillations (GBCO)* as a measure of interactivity (Kidane & Gloor, 2007). This metric, also known as *Rotating Leadership*, sums the number of times individual betweenness centrality scores reach local maxima or minima for a given timeframe (meaning that group members oscillate from more central to more peripheral positions and vice versa). Groups that show higher oscillations in betweenness centrality are usually more dynamically participated and less dominated by static leaders.

We also considered the network size, i.e. the number of social actors, as a metric of interactivity.

The measures of interactivity represent the level of interest and commitment in a discourse. A high interactivity is expressed by a discourse which covers topics of interest for the stakeholders, involving a high number of social actors, who post many tweets, rapidly interact with each other and rotate in conversations – with low average response times, many mentions, answers and retweets (i.e. high nudges) and a high count of betweenness oscillations.

### 3.2.3. Use of language

With regard to the use of language, we calculated the *Sentiment*, *Emotionality* and *Complexity* of the discourse about each core value orientation. Sentiment, computed using a machine learning algorithm included in the software Condor, measures the positivity or the negativity of the language in a range that varies between 0 and 1 – with values below 0.5 representing a negative sentiment and values above a positive one. The algorithms used by Condor to calculate sentiment are detailed in the work of Brönnimann (2013, 2014). Emotionality expresses the variation from neutral sentiment and is equal to 0 if all actors converge towards expressions that are neither positive nor negative; by contrast this measure is high when there is a vivid debate where very positive or negative words emerge strongly (Brönnimann, 2014). Complexity is a proxy for the heterogeneity of the language used. If all actors use common words, with few exceptions, complexity is low. This measure is calculated for each text document by Condor as the probability of the words used to appear in a dictionary (Brönnimann, 2014). In knowledge-sharing contexts where new ideas are important, a higher complexity can be a signal of new knowledge being introduced in the network. While sentiment and emotionality convey the positive (or negative) emotions of stakeholders, complexity can be considered as a proxy of the innovativeness of a discourse, when its values are above the average, but still not too high.

### 3.3. Hierarchy of core values

We suggest determining the level of importance of each core value, based on all the possible combinations of interactivity, connectivity and use of the language metrics. Accordingly, we present, in Table 1, a hierarchy of core values – which has three classes (active, latent and void) adapted from Yoganathan et al. (Yoganathan et al., 2017). It is important to notice that the ratings of "high", "medium" and "low" could also be expressed on comparative bases, depending on company objectives. For example, companies may want to rank their core value statements with respect to their competitors, also taking into consideration time trends. Moreover, not all the metrics included in our framework vary in a predefined range. For example, measures like complexity depend on the context analyzed; therefore, the scores obtained should be analyzed to understand if they are above or below the average, without defining sample-independent thresholds. To this purpose, a min-max normalization of the individual metrics – or a different standardization approach – might come to help.

| Core value | Connectivity metrics | Interactivity metrics | Use of language metrics | Strategy hints |
|---|---|---|---|---|
| *Active* | The measures indicate a range between an intermediate and a high value of connectivity | The measures indicate a high value of interactivity | The measures indicate a positive attitude | *At the heart of any strategic process* |
| *Active but with neutral or negative feelings* | The measures indicate a range between an intermediate and a high value of connectivity | The measures indicate a range between an intermediate and a high value of interactivity | The measures indicate a neutral or negative attitude | *Immediate attention, consider to gradually divest* |
| *Active but on disaggregated groups* | The measures indicate a low value of connectivity | The measures indicate a range between an intermediate and a high value of interactivity | The measures indicate any of the possible attitudes | *Immediate attention, verify the convergence among stakeholders* |
| *Latent* | The measures indicate a range between an intermediate and a high value of connectivity | The measures indicate an intermediate value of interactivity | The measures indicate a neutral or positive attitude | *Periodic attention* |
| *Latent but with negative feelings* | The measures indicate a range between an intermediate and a high value of connectivity | The measures indicate an intermediate value of interactivity | The measures indicate a negative attitude | *Periodic attention, consider to gradually divest* |
| *Latent but on disaggregated groups* | The measures indicate a low value of connectivity | The measures indicate an intermediate value of interactivity | The measures indicate a neutral or positive attitude | *Periodic attention, verify the convergence among stakeholders* |
| *Void* | The measures indicate any of the possible values | The measures indicate a low value of interactivity | The measures indicate any of the possible values | *Consider to gradually divest* |

**Table 1**. Hierarchy of core values

Those values characterized by a very participated dialogue (high number of tweets), frequent interactions among actors, a highly connected network and a shared language, are classified as active. Latent core values are characterized, instead, by a high connectivity and use of language, but also by a smaller level of interaction. Finally, core values are void when the dialogue on them is poor in terms of interactions and use of language, regardless of connectivity. We refer to a 'hierarchy' of core values because active and latent core values are more important for the formulation and implementation of the digital business strategy than those that are void. Actions inspired by active core values are those with the highest probability of achieving a stakeholders-company alignment, which has already proved to positively impact companies performance (Bundy, Vogel, & Zachary, 2018; Henisz, Dorobantu, & Nartey, 2014). Active core values should be at the heart of any strategic process.

Latent core values need periodic attention: they sometimes interest relatively connected actors, but they languish in an area which might be called "contingent disinterest", as their level of interactivity is not yet high. Companies need to nurture stakeholders' interests with an appropriate communication strategy, in order to reaffirm their latent values and strengthen the relationship between corporate identity and the stakeholders' perception of it.

When the level of interactivity is low to non-existent, core values are void. It could be very hard for companies to obtain a positive response from stakeholders when they act following a void values lead. This is the reason why companies should evaluate a process of divestment of void values, especially when low interactivity scores associate with negative feelings. When feelings are neutral, a void classification could also represent outdated core values, as they could be considered as a mere legacy of the past.

Immediate attention is needed for active core values which are perceived negatively and/or associated to disaggregated networks: each case should be carefully examined in order to implement the most appropriate strategy. Companies need to narrowly evaluate the inalienability of those core values that are negatively active, as they could become candidates for the divesting process. They need to rethink themselves, reaffirming their own corporate identity or following stakeholders' pressure for a change. For example, Apple refused to help the FBI to hack the iPhone belonging to one of the killers in the December 2015 mass shooting in San Bernardino, California. Apple chose to remain faithful to its values of protecting the security and the privacy of its customers, despite the pressure of several stakeholders to cooperate with the FBI.

Lastly, a disaggregated network could indicate that the stakeholders' debate takes place in separate groups, with little cross-interaction. It might be the case that stakeholders have different (or conflicting) interests. If the sentiment of these groups alternates between positive and negative, companies should act to recover hostile groups, and examine with attention their topic models to detect reasons for conflicts of interest.

## 4. Results

Figure 1 presents the full network graph, where each twitter user is a node, colored according to the core value orientation which was predominant in his/her tweets.

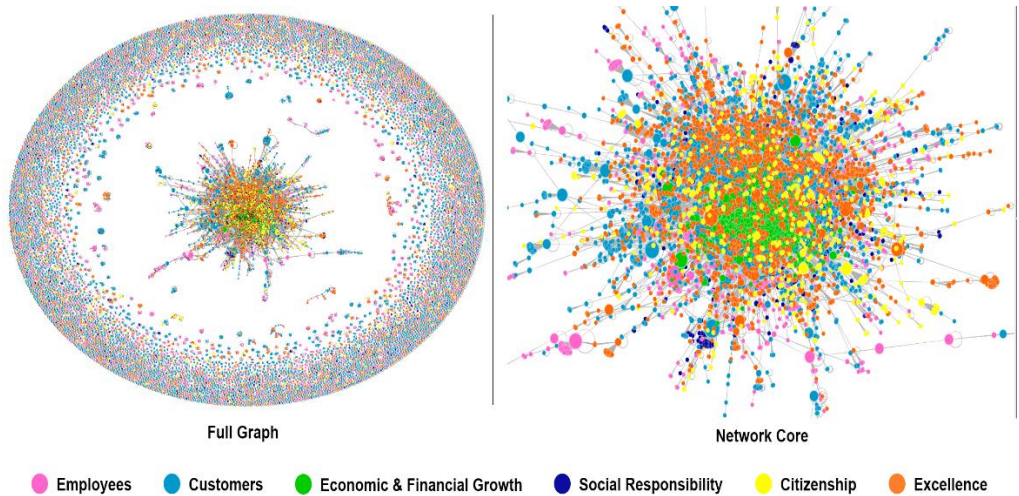

**Figure 1.** Twitter network of the discourse about the core values

We see a large belt of peripheral unconnected nodes around the main network core: these nodes represent users who tweeted without interacting with their peers. The core, on the other hand, is relatively dense and connected, reflecting a vivid debate about the core values. From the picture, we also notice several cross-connections of users, as many of them posted regarding different core values. Indeed, the discourses about distinct core values can be sometimes tightly interrelated.

Table 2 shows the values of the different metrics for each core value orientation. It is important to notice that high scores not always indicate something positive. For example, lower values of Average Response Time (ART) are preferable as they indicate shorter time lapses in the turn taking of social media users. Similarly, very high values of group degree and betweenness centrality indicate a network that is almost totally centralized and has less heterogeneous interaction patterns.

| | | Core Value Orientations | | | | | | | | | | |
|---|---|---|---|---|---|---|---|---|---|---|---|---|
| | | **Customers** | | **Employees** | | **Citizenship** | | **Economic/Financial Growth** | | **Excellence** | | **Social Responsibility** | |
| | **Variables** | **Score** | **MM** | **Score** | **MM** | **Score** | **MM** | **Score** | **MM** | **Score** | **MM** | **Score** | **MM** |
| **Connectivity** | **Group Degree Centrality** | 0.015 | 0.04 | 0.024 | 0.20 | 0.037 | 0.43 | 0.056 | 0.77 | 0.013 | 0.00 | 0.069 | 1.00 |
| | **Group Betweenness Centrality** | 0.016 | 0.08 | 0.024 | 0.19 | 0.063 | 0.74 | 0.01 | 0.00 | 0.029 | 0.26 | 0.082 | 1.00 |
| | **Density** | 0.0001 | 0.00 | 0.0002 | 0.14 | 0.0004 | 0.43 | 0.0006 | 0.71 | 0.0002 | 0.14 | 0.0008 | 1.00 |
| **Interactivity** | **ART** | 5.674 | 1.00 | 3.402 | 0.22 | 2.75 | 0.00 | 2.884 | 0.05 | 4.136 | 0.47 | 4.009 | 0.43 |
| | **Nudges** | 1.073 | 1.00 | 1.055 | 0.75 | 1.065 | 0.89 | 1 | 0.00 | 1.024 | 0.33 | 1 | 0.00 |
| | **Number of Actors** | 12070 | 1.00 | 8041 | 0.61 | 3270 | 0.14 | 2105 | 0.03 | 7872 | 0.59 | 1809 | 0.00 |
| | **Activity** | 19153 | 1.00 | 10957 | 0.50 | 4239 | 0.10 | 2812 | 0.01 | 13437 | 0.65 | 2604 | 0.00 |
| | **Average Activity per Actor** | 1.587 | 0.71 | 1.363 | 0.16 | 1.296 | 0.00 | 1.336 | 0.10 | 1.707 | 1.00 | 1.439 | 0.35 |
| | **Rotating leadership** | 26 | 1.00 | 20 | 0.00 | 23 | 0.50 | 25 | 0.83 | 26 | 1.00 | 21 | 0.17 |
| **Use of language** | **Sentiment** | 0.695 | 0.52 | 0.63 | 0.00 | 0.688 | 0.47 | 0.719 | 0.72 | 0.754 | 1.00 | 0.664 | 0.27 |
| | **Emotionality** | 0.283 | 0.11 | 0.29 | 0.30 | 0.312 | 0.89 | 0.316 | 1.00 | 0.279 | 0.00 | 0.285 | 0.16 |
| | **Complexity** | 7.67 | 0.61 | 7.828 | 1.00 | 7.479 | 0.13 | 7.712 | 0.71 | 7.686 | 0.65 | 7.426 | 0.00 |

*Note.* MM= Min-max normalization of the score.

**Table 2.** Network and semantic metrics of core value orientations.

In our study, users were mostly concerned with Customers orientation, paying a lot of attention to post-sale services and customer assistance: this is the orientation with the highest level of activity. Excellence was the second most tweeted orientation followed by Employees.

Sentiment was on average positive for all the orientations, even if conversations about Economic & Financial Growth and Citizenship were more emotional. The language used was mostly shared among the users, without reaching high peaks of complexity. This measure is higher for the Employees orientation, as many different aspects of the employees' life within a company were discussed (salaries, reductions of the workforce, benefits and facilities, etc.) Figure 2 shows the social network of each orientation.

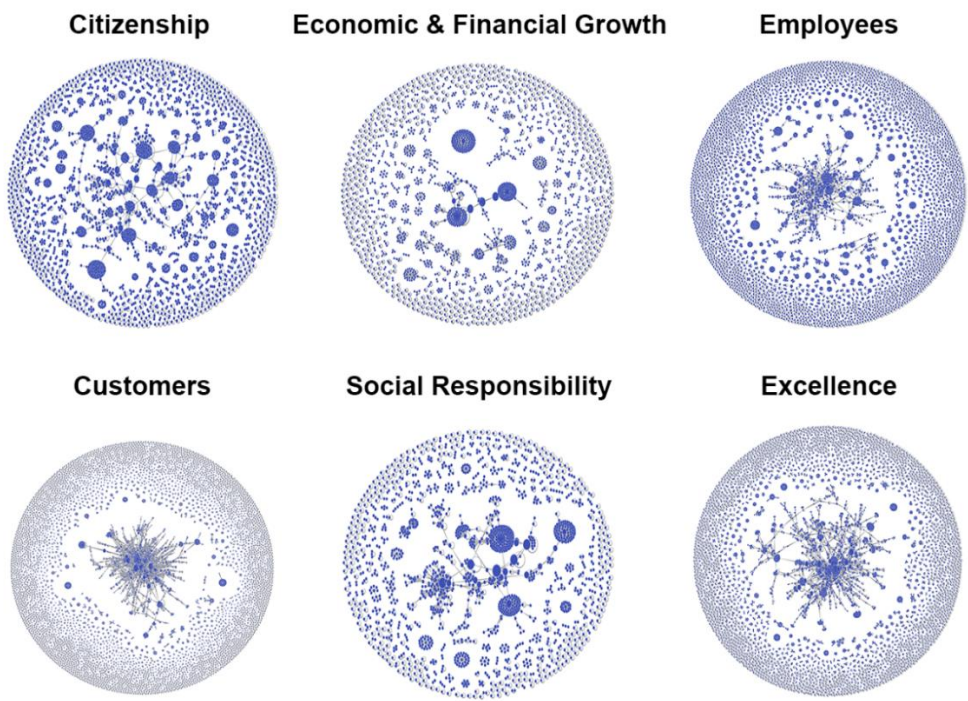

**Figure 2.** Network graphs of the core values orientations

In general, from a more dense and connected network core and from a smaller group of isolated nodes around the core, we could infer a more vivid and participated debate. However, in our results, density is rather low for all the core value orientations, proving the existence of many fragmented voices and the need for firms to foster a more interconnected discourse on Twitter.

Even if Citizenship is an orientation with less tweets than Customers and Employees, coming from a smaller number of actors, the interactions are fast and relatively committed (low response times and higher nudges) and dominated by a set of conversation leaders (each one dealing with a part of the network). Customers, Excellence and Employees seem to be the orientations with the most vivid debate coming both from many individual isolated voices and from users interacting in the network core. Rotating leadership is higher for Customers and Excellence, with low values of network centralization, indicating a more heterogeneous participation and a less dominated discourse. Being many actors in these orientations probably determined an increase in the average of the response times. By contrast, Economic & Financial Growth and Social Responsibility were less attractive orientations for the Twitter users: their network is more centralized (i.e. dominated by fewer voices), smaller, and with a lower level of commitment (less nudges).

Citizenship had almost two times the tweets of Social Responsibility in our sample. This could represent the idea that in some cases companies are expected to act fairly and respect the law, but they are less perceived as subjects responsible for promoting social and environmental change. In other words, there will probably be a higher proportion of stakeholders who think that 'Company X should not pollute the air we breathe', than the proportion of those who think 'Company X should clean the air polluted by somebody else'.

Combining all the results, Table 3 shows the hierarchy of core value orientations as presented in Section 3. We find that for our case study the most active core value orientation is Customers, followed by Employees and Excellence. Even if there is a fairly high participation of social media users around this three core value orientations – and their interactivity is high – the interaction is for a large part produced by dyads or disaggregated groups. Economic & Financial Growth, Citizenship and Social Responsibility, on the other hand, emerge as latent orientations which need periodic attention as they might become active in the future.

| Orientation | Connectivity | Interactivity | Use of language | Classification |
|---|---|---|---|---|
| **Customers** | Low | High | Positive attitude | Active on disaggregated groups |
| **Employees** | Low | Intermediate-High | Positive attitude with high complexity | Active on disaggregated groups |
| **Economic & Financial Growth** | Intermediate-Low | Intermediate-Low | Positive attitude, with fairly high emotionality and complexity | Latent on disaggregated groups |
| **Citizenship** | Intermediate | Intermediate | Positive attitude | Latent |
| **Excellence** | Low | High | Very positive attitude | Active on disaggregated groups |
| **Social Responsibility** | Intermediate | Low | Positive attitude | Latent |

**Table 3.** Hierarchy of core value orientations for the Italian scenario

## 5. Discussion and Conclusions

To face the ever-increasing complexity of the business environment, complexity tools have to be integrated into strategy formulation and implementation processes. Leveraging on big data, the implementation of a digital business strategy may allow the analysis of the complex relationship between firms and stakeholders in the light of core values. In this paper, we propose a methodology for an assessment of the stakeholders' attitude toward a company's core values based on big data techniques, in order to define the desirability and the impact on business strategies of each core value over time. In order to test the proposed methodology, we analyzed two months of Twitter discourse about the core values of the World's most admired companies in Fortune's ranking, from 2013 to 2017. Twitter proved to be a remarkable source of data to study the stakeholders' attitudes towards core values: the discourse we could analyze was significantly participated and involved a relatively high number of social actors. The results showed the different stakeholders' attitude towards core values; therefore, it was possible to build a core value hierarchy, useful to correctly address the strategy formulation and implementation processes. In the Italian scenario, active core value orientations are Customers, Excellence and Employees. Even if the majority of users resulted in isolated nodes or dyads, all these orientations exhibited a relatively connected central core. Such a connected network component allows the identification of network hubs, which is useful when firms want to identify influential nodes and plan a communication strategy aligned with their core values in these three orientations. Consistently, companies (at least those included in our case study) should manage their strategy formulation and implementation processes giving priority to core values which belong to the orientations of Customers, Excellence and Employees. Among these three, Employees is the most critical because it has the lowest (but still positive) sentiment and the highest complexity (indicating different points of view). This probably happens because actions directed at employees are sometimes regarded with suspicion: companies should carefully support their own initiatives with an appropriate communication strategy, in order to emphasize the positive aspects of their actions.

Companies should not neglect those core values which are latent, i.e. belonging to Economic/Financial Growth, Citizenship and Social Responsibility orientations. Firms should consider to fairly distribute resources among strategic activities according to our results, and, at the same time - especially when they consider these values very important for their own corporate identity - they should work to involve more stakeholders. Surprisingly, our findings point out a relative low degree of interest in Corporate Social Responsibility (CSR, i.e. Citizenship and Social Responsibility orientations together); however, many past studies considered CSR to be a strategic necessity (Babiak & Trendafilova, 2011; Barchiesi, Castellan, & Costa, 2016; Barchiesi & La Bella, 2014; Calabrese, Costa, Menichini, Rosati, & Sanfelice, 2013; Falkenberg & Brunsæl, 2011; Hsu, 2018). This means that the debate on CSR importance is far from finding a unique interpretation and that stakeholders' attitudes toward CSR should be deeper examined in further studies (Bhardwaj, Chatterjee, Demir, & Turut, 2018). Moreover, the fact that Citizenship had almost twice the tweets of Social Responsibility and a higher interactivity shows a clear distinction between these two core value orientations. Stakeholders expect companies to obey the law, but are much less concerned with company contribution to the improvement of society and the environment, while, several studies on CSR emphasize the aspect of "voluntariness" of firm commitment, i.e. they maintain the high importance of going beyond what is prescribed by the law (e.g., Piacentini, MacFadyen, & Eadie, 2000; *Promoting a European Framework for Corporate Social Responsibilities*, 2001).

Our study contributes to the literature on the business use of complexity tools, showing how big data extracted from social media can be used to support strategy formulation and implementation processes (Merendino et al., 2018; Sivarajah, Kamal, Irani, & Weerakkody, 2017). Our study also adds to the research on honest signals (Pentland, 2008), and presents a new field of application for the metrics framed by Gloor and colleagues (P. A. Gloor, 2017; P Gloor, 2017; Peter Gloor et al., 2017).

The methodology we propose can be used by company managers to understand the positioning of their company's core values, also with respect to competitors. The attitudes of stakeholders could be compared across different social media, such as Facebook, Reddit and online forums. Accordingly, managers could use our approach to see whether their core values classify as active, latent or void; starting from the strategy hints presented in Table 1, they could arrange the most appropriate corrective actions – such as rethinking their company's values or better supporting the existing ones with new communication campaigns. Managerial decisions are highly dependent on the strategic orientation that a company follows (Mikalef, Pateli, Batenburg, & Wetering, 2015). Our approach supports decision-making and strategy-formulation in the era of big data, which demand cognitive and adaptive capabilities of managers, who are prompted to make high-level decisions more rapidly (Merendino et al., 2018).

Our study has some limitations that originate from the choice of analyzing an Italian scenario for a limited timeframe (two months), with also a possible influence of the Italian culture over the attitude of Twitter users. The results we obtained are limited to our case study and might change in different settings. Accordingly, we advocate future research to test our methodology considering, for example, other companies, different countries and cultural settings, or more extended timeframes. Our methodology can be replicated by companies of any size, given that their core values are properly stated. This could lead to different results, especially in the case of special events taking place during the time of analysis. For example, a latent Social Responsibility core value could become active because of an environmental disaster. This is aligned with our approach, meant to provide a classification of core values which can change over time. Our methodology also allows managers to detect priorities while triggering contingency strategies.

Lastly, the framework of analysis that we propose in this study is not limited to Twitter and could also be applied to data extracted from other social media platforms, such as Facebook. A future development of our work could consider the integration of the measures framed by Gloor et al. (Peter Gloor et al., 2017) with recent tools and metrics designed for the advanced assessment of brand importance, through semantic and social network analysis. For example, the Semantic Brand Score, a novel measure of brand importance, could further help the ranking of core values, according to their relevance and embeddedness in the social media discourse (Fronzetti Colladon, 2018).

**References**

Agle, R. B., & Caldwell, C. (1999). Understanding research on values in business. *Business and Society*, *38*(3), 326.

Akaah, I. P., & Lund, D. (1994). The influence of personal and organizational values on marketing professionals' ethical behavior. *Journal of Business Ethics*, *13*(6), 417–430. https://doi.org/10.1007/BF00881450

Arias, M., Arratia, A., & Xuriguera, R. (2014). Forecasting with Twitter Data. *ACM*

*Transactions on Intelligent Systems and Technology*, *5*(1), 1–24. https://doi.org/10.1145/2542182.2542190

Babiak, K., & Trendafilova, S. (2011). CSR and environmental responsibility: Motives and pressures to adopt green management practices. *Corporate Social Responsibility and Environmental Management*, *18*(1), 11–24. https://doi.org/10.1002/csr.229

Banker, R. D., Hu, N., Luftman, J., & Pavlou, P. A. (2010). CIO Reporting Structure, Strategic Positioning, and Firm Performance: To Whom Should the CIO Report? *SSRN Electronic Journal*, *35*(2), 487–504. https://doi.org/10.2139/ssrn.1557874

Barchiesi, M. A., Castellan, S., & Costa, R. (2016). In the eye of the beholder: Communicating CSR through color in packaging design. *Journal of Marketing Communications*, *7266*(August), 1–14. https://doi.org/10.1080/13527266.2016.1224771

Barchiesi, M. A., & La Bella, A. (2014). An Analysis of the Organizational Core Values of the World's Most Admired Companies. *Knowledge and Process Management*, *21*(3), 159–166. https://doi.org/10.1002/kpm.1447

Beer, M., & Eisenstat, R. A. (2000). The Silent Killers of Strategy Implementation and Learning. *Sloan Management Review*, *41*(4), 29–40.

Bharadwaj, A., El Sawy, O. a., Pavlou, P. a., & Venkatraman, N. (2013). Digital Business Strategy: Toward a Next Generation of Insights. *MIS Quarterly*, *37*(2), 471–482. https://doi.org/10.1.1.216.1018

Bhardwaj, P., Chatterjee, P., Demir, K. D., & Turut, O. (2018). When and how is corporate social responsibility profitable? *Journal of Business Research*, *84*, 206–219.

https://doi.org/10.1016/j.jbusres.2017.11.026

Brönnimann, L. (2013). *Multilanguage sentiment-analysis of Twitter data on the example of Swiss politicians*. Windisch, Switzerland. Retrieved from http://www.twitterpolitiker.ch/Paper_Swiss_Politicians_On_Twitter.pdf

Brönnimann, L. (2014). *Analyse der Verbreitung von Innovationen in sozialen Netzwerken*. Retrieved from http://www.twitterpolitiker.ch/documents/Master_Thesis_Lucas_Broennimann.pdf

Bryson, J. M., Crosby, B. C., & Bryson, J. K. (2009). Understanding Strategic Planning and the Formulation and Implementation of Strategic Plans as a Way of Knowing: The Contributions of Actor-Network Theory. *International Public Management Journal*, *12*(2), 172–207. https://doi.org/10.1080/10967490902873473

Bundy, J., Vogel, R. M., & Zachary, M. A. (2018). Organization–stakeholder fit: A dynamic theory of cooperation, compromise, and conflict between an organization and its stakeholders. *Strategic Management Journal*, *39*(2), 476–501. https://doi.org/10.1002/smj.2736

Calabrese, A., Costa, R., Menichini, T., Rosati, F., & Sanfelice, G. (2013). Turning corporate social responsibility-driven opportunities in competitive advantages: A two-dimensional model. *Knowledge and Process Management*, *20*(1), 50–58. https://doi.org/10.1002/kpm.1401

Chatman, J. A. (1989). Improving Interactional Organizational Research: A Model of Person-Organization Fit. *Academy of Management Review*, *14*(3), 333–349. https://doi.org/10.5465/AMR.1989.4279063

Cording, M., Harrison, J. S., Hoskisson, R. E., & Jonsen, K. (2014). Walking the Talk: A Multistakeholder Exploration of Organizational Authenticity, Employee Productivity, and Post-Merger Performance. *Academy of Management Perspectives*, *28*(1), 38–56. https://doi.org/10.5465/amp.2013.0002

Culnan, M. J., McHugh, P. J., & Zubillaga, J. I. (2010). How Large U.S. Companies Can Use Twitter and Other Social Media to Gain Business Value. *MIS Quarterly Executive*, *9*(4), 243–259. https://doi.org/10.1108/07378830510636300

De Mauro, A., Greco, M., & Grimaldi, M. (2016). A formal definition of Big Data based on its essential features. *Library Review*, *65*(3), 122–135. https://doi.org/10.1108/LR-06-2015-0061

Ettlie, J. E., & Pavlou, P. A. (2006). Technology-based new product development partnerships. *Decision Sciences*, *37*(2), 117–147. https://doi.org/10.1111/j.1540-5915.2006.00119.x

Falkenberg, J., & Brunsæl, P. (2011). Corporate Social Responsibility: A Strategic Advantage or a Strategic Necessity? *Journal of Business Ethics*, *99*(SUPPL. 1), 9–16. https://doi.org/10.1007/s10551-011-1161-x

Felix, R., Rauschnabel, P. A., & Hinsch, C. (2017). Elements of strategic social media marketing: A holistic framework. *Journal of Business Research*, *70*, 118–126. https://doi.org/10.1016/j.jbusres.2016.05.001

Ferraro, G., & Iovanella, A. (2015). Organizing Collaboration in Inter-Organizational Innovation Networks, from Orchestration to Choreography. *International Journal of Engineering Business Management*, *7*, 24. https://doi.org/10.5772/61802

Ferraro, G., Iovanella, A., & Cinelli, M. (2017). Network processes for collaborative innovation. *Search Results International Journal of Entrepreneurship and Small Business*, in press.

Fitzgerald, G. A., & Desjardins, N. M. (2004). Organizational Values and Their Relation to Organizational Performance Outcomes. *Atlantic Journal of Communication*. https://doi.org/10.1207/s15456889ajc1203_1

Fronzetti Colladon, A. (2018). The Semantic Brand Score. *Journal of Business Research*, *88*, 150–160. https://doi.org/10.1016/j.jbusres.2018.03.026

Gimeno, A., Labadie, G. J., Saris, W., & Mendoza, X. (2006). Internal factors of family business performance: an integrated theoretical model. In *Handbook of research on family business* (pp. 145–164).

Gloor, P. A. (2017). *Sociometrics and Human Relationships: Analyzing Social Networks to Manage Brands, Predict Trends, and Improve Organizational Performance*. London, UK: Emerald Publishing Limited.

Gloor, P. A., Fronzetti Colladon, A., Miller, C. Z., & Pellegrini, R. (2016). Measuring the level of global awareness on social media. In M. Zylka, H. Führes, A. Fronzetti Colladon, & P. A. Gloor (Eds.), *Designing Networks for Innovation and Improvisation* (pp. 125–139). Cham, Switzerland: Springer International Publishing. https://doi.org/10.1007/978-3-319-42697-6_13

Gloor, P. (2017). The Signal Layer: Six Honest Signals of Collaboration. In *Swarm Leadership and the Collective Mind* (pp. 91–104). Bingley, UK: Emerald Publishing Limited. https://doi.org/10.1108/978-1-78714-200-820171006

Gloor, Peter, Fronzetti Colladon, A., Giacomelli, G., Saran, T., & Grippa, F. (2017). The impact of virtual mirroring on customer satisfaction. *Journal of Business Research*, *75*, 67–76. https://doi.org/10.1016/j.jbusres.2017.02.010

Gupta, M., & George, J. F. (2016). Toward the development of a big data analytics capability. *Information and Management*, *53*(8), 1049–1064. https://doi.org/10.1016/j.im.2016.07.004

Henisz, W. J., Dorobantu, S., & Nartey, L. J. (2014). Spinning gold: The financial returns to stakeholder engagement. *Strategic Management Journal*. https://doi.org/10.1002/smj.2180

Higgins, J. M., & Mcallaster, C. (2004). If you Want StrategicChange, Don't Forget to Change your Cultural Aartifacts. *Journal of Change Management*, *4*(1), 63–73. https://doi.org/10.1080/1469701032000154926

Hitlin, S., & Piliavin, J. A. (2004). Values: Reviving a Dormant Concept. *Annual Review of Sociology*, *30*(1), 359–393. https://doi.org/10.1146/annurev.soc.30.012703.110640

Hsu, F. J. (2018). Does corporate social responsibility extend firm life-cycles? *Management Decision*, MD-09-2017-0865. https://doi.org/10.1108/MD-09-2017-0865

Ind, N. (2004). Living the brand: How to transform every member of your organization into a brand ambassador. *Kogan Page, London*. https://doi.org/10.1108/17538330910942799

Janssen, M., van der Voort, H., & Wahyudi, A. (2017). Factors influencing big data decision-making quality. *Journal of Business Research*, *70*, 338–345. https://doi.org/10.1016/j.jbusres.2016.08.007

Kidane, Y. H., & Gloor, P. A. (2007). Correlating temporal communication patterns of the Eclipse open source community with performance and creativity. *Computational and*

*Mathematical Organization Theory*, *13*(1), 17–27.

Kohli, R., & Grover, V. (2008). Business Value of IT : An Essay on Expanding Research Directions to Keep up with the Times. *Journal of the Association Fo Information Systems*, *9*(1), 23–39. https://doi.org/Article

Lovejoy, K., Waters, R. D., & Saxton, G. D. (2012). Engaging stakeholders through Twitter: How nonprofit organizations are getting more out of 140 characters or less. *Public Relations Review*, *38*(2), 313–318.

Mamic, L. I., & Almaraz, I. A. (2013). How the Larger Corporations Engage with Stakeholders through Twitter. *International Journal of Market Research*, *55*(6), 851–872. https://doi.org/10.2501/IJMR-2013-070

Merendino, A., Dibb, S., Meadows, M., Quinn, L., Wilson, D., Simkin, L., & Canhoto, A. (2018). Big data, big decisions: The impact of big data on board level decision-making. *Journal of Business Research*, *93*, 67–78. https://doi.org/10.1016/j.jbusres.2018.08.029

Mikalef, P., Framnes, V. A., Danielsen, F., Krogstie, J., & Olsen, D. H. (2017). Big Data Analytics Capability : Antecedents and Business Value. In *Twenty First Pacific Asia Conference on Information Systems*.

Mikalef, P., Ilias, P., Giannakos, M., Krogstie, J., & Lekakos, G. (2016). Big Data and Strategy: A research Framework. *MCIS 2016 Proceedings*.

Mikalef, P., Pateli, A., Batenburg, R. S., & Wetering, R. van de. (2015). Purchasing alignment under multiple contingencies: a configuration theory approach. *Industrial Management & Data Systems*, *115*(4), 625–645. https://doi.org/10.1108/IMDS-10-2014-0298

Ngai, E. W. T., Xiu, L., & Chau, D. C. K. (2009). Application of data mining techniques in customer relationship management: A literature review and classification. *Expert Systems with Applications*, *36*(2 PART 2), 2592–2602. https://doi.org/10.1016/j.eswa.2008.02.021

Pappas, I. O., Mikalef, P., Giannakos, M. N., Krogstie, J., & Lekakos, G. (2017). *Social media and analytics for competitive performance: A conceptual research framework*. *Lecture Notes in Business Information Processing* (Vol. 263). https://doi.org/10.1007/978-3-319-52464-1_19

Parada, M. J., Nordqvist, M., & Gimeno, A. (2010). Institutionalizing the Family Business: The Role of Professional Associations in Fostering a Change of Values. *Family Business Review*, *23*(4), 355–372. https://doi.org/10.1177/0894486510381756

Pavlou, P. A., & El Sawy, O. A. (2006). From IT leveraging competence to competitive advantage in turbulent environments: The case of new product development. *Information Systems Research*, *17*(3), 198–227. https://doi.org/10.1287/isre.1060.0094

Pekka, A. (2010). Social media, reputation risk and ambient publicity management. *Strategy & Leadership*, *38*(6), 43–49. https://doi.org/10.1108/10878571011088069

Pentland, A. (2008). *Honest Signals*. Cambridge, MA: MIT Press.

Piacentini, M., MacFadyen, L., & Eadie, D. (2000). Corporate social responsibility in food retailing. *International Journal of Retail & Distribution Management*, *28*(11), 459–469. https://doi.org/10.1108/09590550010356822

*Promoting a European Framework for Corporate Social Responsibilities*. (2001). Brussels. Retrieved from

https://www.google.com/url?sa=t&rct=j&q=&esrc=s&source=web&cd=1&ved=2ahUKEwj
NwPjT4rnkAhVFzaQKHfEyDVUQFjAAegQIAhAC&url=http%3A%2F%2Feuropa.eu%2
Frapid%2Fpress-release_DOC-01-9_en.pdf&usg=AOvVaw3SIK0UIX4-K4XVJXxz2ZUk

Pryor, M. G., Anderson, D., Toombs, L. A., & Humphreys, J. H. (2007). Strategic
Implementation as a Core Competency The 5P's Model. *Journal of Management Research*,
*7.1*, 3–17.

Rybalko, S., & Seltzer, T. (2010). Dialogic communication in 140 characters or less: How
Fortune 500 companies engage stakeholders using Twitter. *Public Relations Review*, *36*(4),
336–341. https://doi.org/10.1016/j.pubrev.2010.08.004

Sambamurthy, V., Bharadwaj, A., & Grover, V. (2003). Shaping Agility Through Digital
Options: Reconceptualizing the Role of Information Technology in Contemporary Firms.
*MIS Quarterly*, *27*(2), 237–263. https://doi.org/10.2307/30036530

Schildt, H. (2017). Big data and organizational design–the brave new world of algorithmic
management and computer augmented transparency. *Innovation: Management, Policy and
Practice*, *19*(1), 23–30. https://doi.org/10.1080/14479338.2016.1252043

Sedereviciute, K., & Valentini, C. (2011). Towards a More Holistic Stakeholder Analysis
Approach. Mapping Known and Undiscovered Stakeholders from Social Media.
*International Journal of Strategic Communication*, *5*(4), 221–239.
https://doi.org/10.1080/1553118X.2011.592170

Singhapakdi, A., & Vitell, S. J. (1993). Personal and professional values underlying the ethical
judgments of marketers. *Journal of Business Ethics*, *12*(7), 525–533.
https://doi.org/10.1007/BF00872374

Sivarajah, U., Kamal, M. M., Irani, Z., & Weerakkody, V. (2017). Critical analysis of Big Data challenges and analytical methods. *Journal of Business Research*, *70*, 263–286. https://doi.org/10.1016/j.jbusres.2016.08.001

Tajvidi, M., Richard, M.-O., Wang, Y., & Hajli, N. (2018). Brand co-creation through social commerce information sharing: The role of social media. *Journal of Business Research*. https://doi.org/10.1016/j.jbusres.2018.06.008

Urde, M. (2009). Uncovering the corporate brand's core values. *Management Decision*, *47*(4), 616–638. https://doi.org/10.1108/00251740910959459

Vandenberghe, C., & Peiro, J. M. (1999). Organizational and Individual Values: Their Main and Combined Effects on Work Attitudes and Perceptions. *European Journal of Work and Organizational Psychology*, *8*(4), 569–581. https://doi.org/10.1080/135943299398177

Wang, Y., Kung, L., Wang, W. Y. C., & Cegielski, C. G. (2017). An integrated big data analytics-enabled transformation model: Application to health care. *Information & Management*. https://doi.org/10.1016/j.im.2017.04.001

Wasserman, S., & Faust, K. (1994). *Social Network Analysis: Methods and Applications*. New York, NY: Cambridge University Press. https://doi.org/10.1525/ae.1997.24.1.219

Waterman, R. (2007). *In Search of Excellence. Bloomsbury Business Library - Management Library, 2007, p40-40-40*.

Williams, S. L. (2002). Strategic planning and organizational values: links to alignment. *Human Resource Development International*, *5*(2), 217–233. https://doi.org/10.1080/13678860110057638

Xie, K., Wu, Y., Xiao, J., & Hu, Q. (2016). Value co-creation between firms and customers: The role of big data-based cooperative assets. *Information and Management*, *53*(8), 1034–1048. https://doi.org/10.1016/j.im.2016.06.003

Yaqoob, I., Hashem, I. A. T., Gani, A., Mokhtar, S., Ahmed, E., Anuar, N. B., & Vasilakos, A. V. (2016). Big data: From beginning to future. *International Journal of Information Management*. https://doi.org/10.1016/j.ijinfomgt.2016.07.009

Yoganathan, V., McLeay, F., Osburg, V.-S., & Hart, D. (2017). The Core Value Compass: visually evaluating the goodness of brands that do good. *Journal of Brand Management*. https://doi.org/10.1057/s41262-017-0074-0

Zhang, X., Fuehres, H., & Gloor, P. A. (2011). Predicting Stock Market Indicators Through Twitter "I hope it is not as bad as I fear." *Procedia - Social and Behavioral Sciences*, *26*, 55–62. https://doi.org/10.1016/j.sbspro.2011.10.562